\documentclass[runningheads, oribibl]{llncs}

\usepackage{amsmath, amssymb,amscd}
\usepackage{graphicx}
\usepackage{url}

\newcommand{\keywords}[1]{\par\addvspace\baselineskip
\noindent\keywordname\enspace\ignorespaces#1}

\begin{document}
\title{Shape-based peak identification for ChIP-Seq}
\titlerunning{Shape-based peak identification for ChIP-Seq}
\author{Valerie Hower\and Steven N. Evans \and Lior Pachter}
\institute{University of California, Berkeley}
\maketitle
\begin{abstract}
We present a new algorithm for the identification of bound regions from ChIP-seq experiments. Our method for identifying statistically significant peaks from read coverage is inspired by the notion of persistence in topological data analysis and provides a non-parametric approach that is robust to noise in experiments. Specifically, our method reduces the peak calling problem to the study of tree-based statistics derived from the data. We demonstrate the accuracy of our method on existing datasets, and we show that it can discover previously missed regions and can more clearly discriminate between multiple binding events. The software T-PIC (\underline{T}ree shape \underline{P}eak \underline{I}dentification for \underline{C}hIP-Seq) is available at 

\url{http://math.berkeley.edu/~vhower/tpic.html}

\keywords{ChIP-Seq, peak calling, topological data analysis}
\end{abstract}
\section{Introduction}
With rapidly decreasing costs of sequencing, next-generation sequencing assays are increasingly being used for molecular measurements  \cite{Wold2008}.  These techniques generate millions of short reads and massive data sets, making it computationally challenging to properly analyze the data.  One such assay, called ChIP-Seq (chromatin immunoprecipitation followed by sequencing), is used to determine DNA binding sites of a protein (see \cite{Barski2009,Park2009} for a review).  In ChIP-Seq, protein is first cross-linked to DNA and the fragments subsequently sheared. Following a size selection step that enriches for fragments of specified lengths, the fragments ends are sequenced, and the resulting reads are aligned to the genome.  Reads pile up at bound regions, but due to the ``noise'' inherent in the assay,  calling ``peaks'' is not a straightforward task.  

While there are many current algorithms for analyzing ChIP-Seq data  
\cite{Blahnik2010, Fejes2008, Ji2008,Johnson2007, Jothi2008, Kharchenko2008,Nix2008, Robertson2007,PeakSeq, MACS}, there is still room for improvement as most rely on ad-hoc heuristics including height thresholds and simplistic filters. We present a new approach for calling peaks that uses  not only the height but also the \emph{shape} of a putative peak.  We wish to quantify ``peakness''  of the data while avoiding too many heuristics.  Specifically, we use shape to differentiate between random and nonrandom fragment placement on the genome in a statistically significant way.  We compare our predictions to those made by PeakSeq \cite{PeakSeq} and MACS \cite{MACS} using two published data sets.   
\section{Methods}
\subsection{Overview of the algorithm}
The input to our algorithm consists of the aligned reads for both the sample and input control.  We create a `coverage function'---a map $f$ from the genomic coordinates to the nonnegative integers--- by extending each of the aligned sample reads to the average fragment length $L$.  The `height' $f(t_0)$ at a nucleotide $t_0$ is the number of extended reads that contain $t_0$.  This coverage function is the data that we analyze.  Motivated by current work in topological data analysis (TDA) \cite{Carlsson-2009},  we determine protein binding sites based on the shape of the coverage function.  Our measure of shape comes from associating a rooted tree with each putative peak; a rooted tree is a type of graph that has a vertex specified as the ``root'' and all edges directed away from the root.  This tree captures the important local features of the coverage function (see \cite{EvHoPa} for further details).  We summarize shape using a numerical descriptor of the tree: a \emph{tree shape statistic} $\mathcal{M}$ that differentiates random from nonrandom coverage. 

 Our work uses the theory developed in \cite{EvHoPa} to model random fragment placement on a genome.  This theory yields Galton-Watson trees with generation dependent offspring distributions (see \cite{MR0400434, MR0065835,MR1991122} for background) that describe the shape of the coverage function and depend on the rate of the reads being aligned.  Using the rate $ \frac{\#\; \mathrm{ of\;  reads\; mapped}}{\mathrm{length\; of\; genome}}$ to analyze data on the entire genome would be inappropriate.  In the absence of binding, some genomic regions receive a large number of fragments while others receive very few \cite{Pepke2009}.  Hence, we first divide the genome into regions where we expect the background to be homogeneous across each region and calculate a rate for each region separately.  

 Then, for each region $R$, we obtain a collection of subtrees/possible peaks from the segments in the set $$\mathcal{S}=\{ t\in R \; | \; f(t)\ge h_R\},$$ for a height $h_R$ (a segment is a subset of $\mathcal{S}$ consisting of contiguous nucleotides).   We use segments with at least $10$ base pairs and build trees from the coverage function along each segment.  Next, we simulate $10,000$ random trees with root at height $h_R$, and we record our tree shape statistic $\mathcal{M}$ for each tree.  We use the resulting empirical distribution to obtain an estimate of the function 
$$G_{R}(m):= \mathbb{P}\{\mathcal{M}(T) \ge m \; | \; T \mbox{ is a random tree from region }R\}.$$

\begin{figure}
\begin{center}
\includegraphics[width=4.2in]{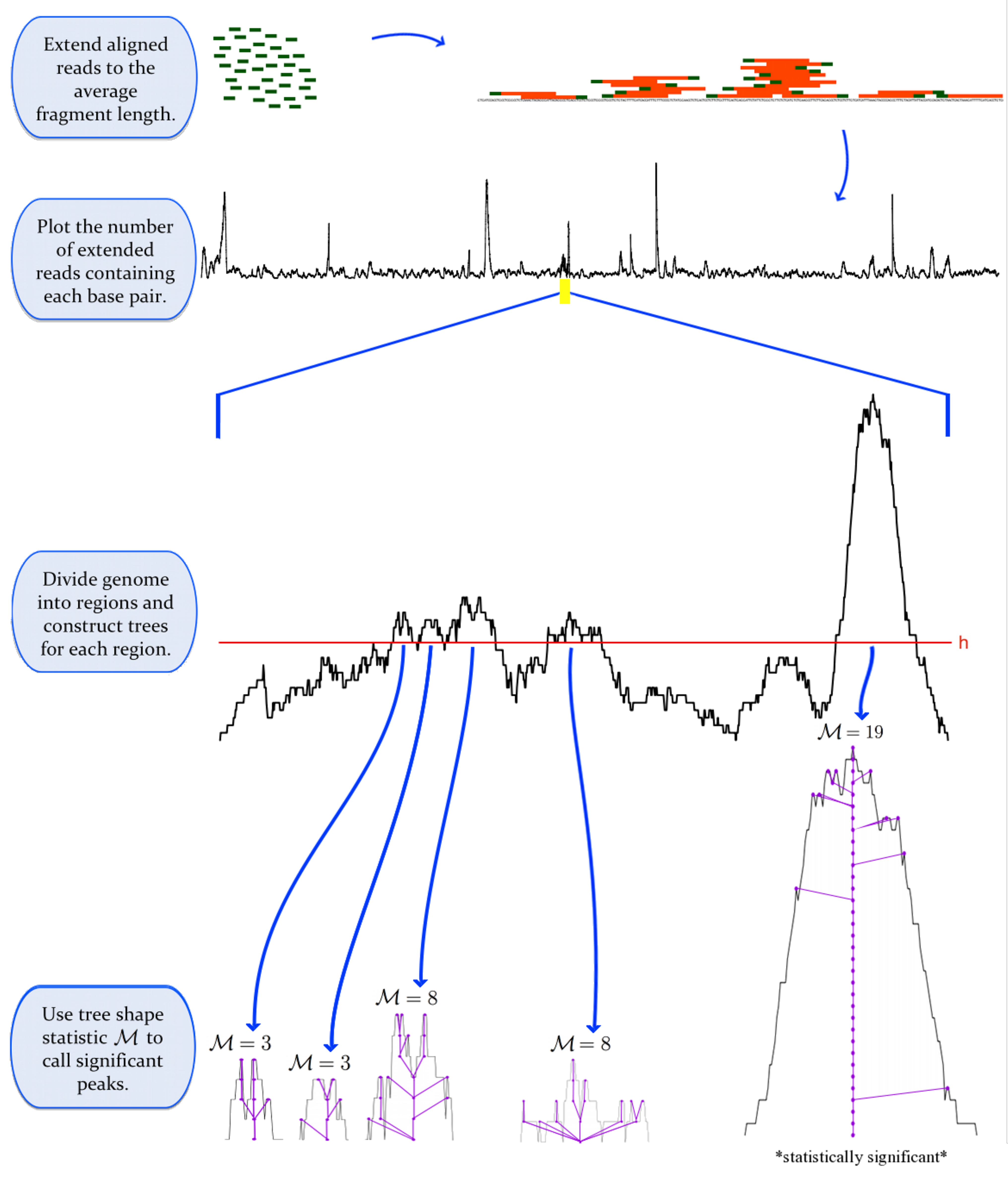}
\caption{In our method, aligned reads are extended to the average fragment length (for single end sequencing), and a coverage function records the number of extended reads containing each base pair.  Trees capturing the shape of the coverage function are constructed and a tree shape statistic measuring the size of a maximal matching $\mathcal{M}$ is computed. By comparison to a null model derived from the expected shape of random trees, significant peaks are identified.}
\label{BuildT}
\end{center}
\end{figure}

 We then compute $\mathcal{M}(T)$ for each tree $T$ in $R$ and assign the probability $G_{R}(\mathcal{M}(T))$ to $T$.    Once we have assigned probabilities to all subtrees obtained from the data, we account for multiple hypothesis testing using a Benjamini--Hochberg correction \cite{Bejamini}.  The significant trees are called as peaks, and we merge two called peaks in bordering regions provided the gap between them is less than $L$.  Figure \ref{BuildT} gives a pictorial sketch of our method.  

 \subsection{The shape of random fragment placement} 
\begin{figure}
\begin{center}
\includegraphics[width=3in]{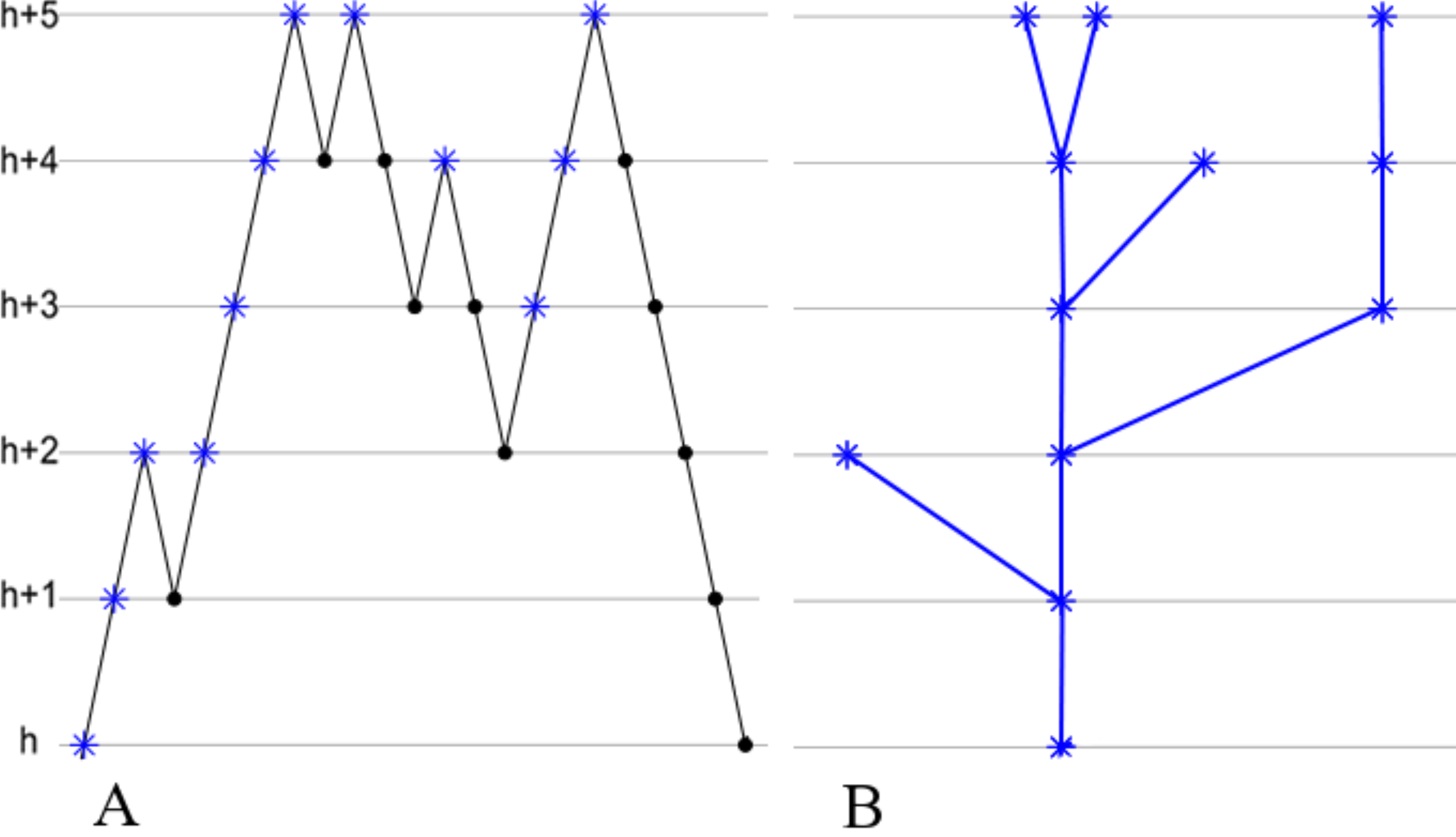}
\caption{An example of lattice path excursion (A) and its associated rooted tree (B) is given.  The rooted tree is obtained by taking equivalence classes of vertices in (A), as explained in  \cite{EvHoPa, Evans-book}.  The vertices in (A) that are chosen representatives for the equivalence classes are depicted with blue stars.     }
\label{jumptree}
\end{center}
\end{figure}

As illustrated in Figure \ref{jumptree} and discussed in \cite{EvHoPa, Evans-book}, there is an equivalence between lattice path excursions (starting and ending at height $h$) and rooted trees with root at height $h$.  We use this correspondence and simulate trees coming from random fragment placement using lattice path excursions for a discrete-time Markov chain $Z$ on the nonnegative integers.

The Markov chain $Z$ approximates the jump skeleton of a coverage function created from random fragment placement \cite{EvHoPa} and has transition probabilities
\begin{equation}\label{markov}
P(i,j) 
=
\begin{cases}
1,& \quad \text{if $i=0$ and $j=1$}, \\
p(i),& \quad \text{if $i \ge 1$ and $j=i+1$}, \\
1 - p(i),& \quad \text{if $i \ge 1$ and $j=i-1$}, \\
0,& \quad \text{otherwise},
\end{cases}
\end{equation}
where 
$$
p(k)= \theta e^{-\theta} \int_0^1w^ke^{\theta w}dw \quad \mbox{for} \quad k\ge 1
$$
and $\theta = \rho L$ is the expected height of the coverage function created with rate $\rho$ and mean fragment length $L$.
We integrate by parts and obtain the recursion 
\begin{eqnarray*}
&p(k)=1-\frac{k}{\theta}p(k-1), \quad k\ge 2, \\
& p(1)=1- \frac{1}{\theta}+\frac{e^{-\theta}}{\theta},
\end{eqnarray*} 
which allows one to solve for $p(k)$.  In our simulation, we first step to height $h+1$ and then either return to height $h$ with probability $1-p(h+1)$ or step to height $h+2$ with probability $p(h+1)$.  We continue visiting states of the Markov chain---jumping up or down at state $i$ depending on the probability $p(i)$---until we return to state $h$.  This lattice path excursion corresponds to a simulated tree with root at height $h$.

\subsection{A tree shape statistic $\mathcal{M}$ to measure ``peakness''}
\begin{figure}
\begin{center}
\includegraphics[width=3.3in]{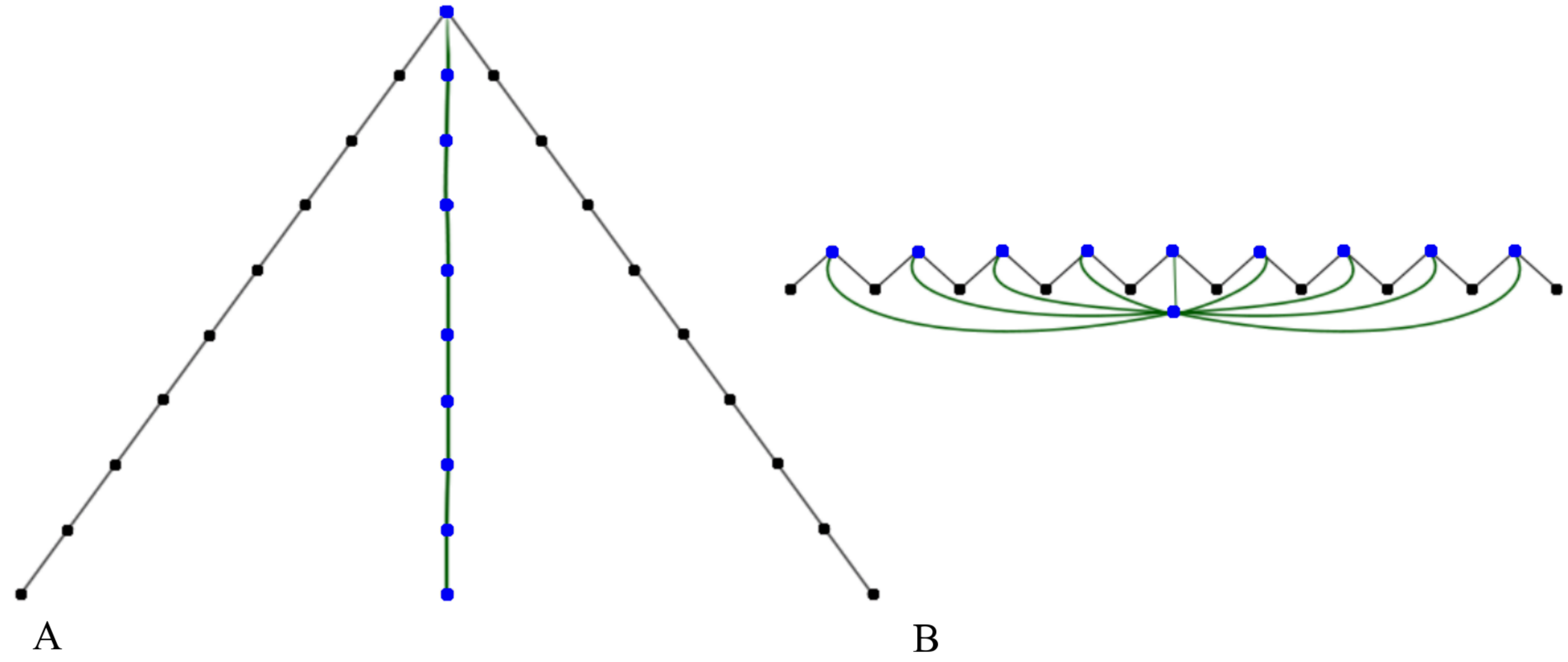}
\caption{Two extremal trees are represented---the path $P_{10}$ (A) and star graph $S_{10}$ (B) on $10$ vertices (blue vertices and green edges)---together with the jump skeleta (black vertices and edges) that give rise to the trees. }
\label{extreme}
\end{center}
\end{figure}

In order to select an appropriate tree statistic, we consider the extremes.  Figure \ref{extreme} depicts the jump skeletons and corresponding rooted trees for, respectively, a perfect peak and perfect noise.  We select a tree statistic $\mathcal{M}$ that attains its extremal values on the path $P_n$ and the star graph $S_n$ on $n$ vertices.  

A \emph{matching} of a tree $T$ is a subset $M$ of edges of $T$ so that no two edges in $M$ share a common vertex of $T$.  A matching $M$ is a \emph{maximal matching} provided $|M|\ge |M^{\prime}|$ for any other matching $M^{\prime}$ of $T$.    We define $\mathcal{M}(T)=|M|$ for $M$ a maximal matching of $T$, and we have $\mathcal{M}(T)\le \lfloor \frac{n}{2}\rfloor = \mathcal{M}(P_n)$ and $\mathcal{M}(T)\ge 1=\mathcal{M}(S_n)$ for any tree $T$ with $n$ vertices.  In our implementation, we calculate the tree shape statistic $\mathcal{M}$ using the algorithm in \cite{bhamidi2009}.

\subsection{Subdividing the genome into regions}
We subdivide the genome into regions based on the input control and perform our analysis on each region separately.  Given the input, we calculate a local rate function $$\zeta(t)=\frac{\mbox{\# of input tags originating in } [t-500,t+500]}{1000}.$$  We then discretize $\zeta$ into a step function as follows.  For each chromosome, we begin with the interval $I=[1,K],$ where $K$ is a user specified integer, and find the average of $\zeta$ over $I$.  We extend $I$, adding nucleotides  $K+1, K+2, \cdots , t_0$ until $\zeta(t_0+1)$ differs from the computed average $\zeta$ by more than a  fixed user specified value $D$.  The next interval begins as $[t_0+1, t_0+K]$, and it is extended until $\zeta$ jumps away from its average by more than $D$.  For the human genome, we use $K=10,000$, but one could use a smaller $K$ for shorter genomes.  Additionally, we use $D=5$.  Once all the intervals are determined for all chromosomes, we round each average $\zeta$ to the nearest integer and define (disconnected) regions $R_j$ based on the intervals whose average $\zeta$ rounds to $j$.  We calculate the local rate $$\rho_j=\frac{\#\mbox{ of tags in data originating in }R_j}{\displaystyle{\sum_{\coprod I=R_j} }\mbox{length}(I)}$$ for the data along $R_j$. 
\begin{remark}
In the absence of input control, one could account for some of the rate variability by dividing the genome into intervals of a fixed size and using a rate $\rho$ that depends on the mappability of the interval,
$$\rho:=\frac{\# \mbox{ of tags originating in interval}}{\# \mbox{ of uniquely mappable bases in interval}}.$$
The mappability of an interval can be computed using the methods in \cite{PeakSeqCode}. 
\end{remark}

\subsection{Choosing the height $h_R$ for a region $R$.}
Care must be used when selecting $h_R$, which depends on the rate calculated for region $R$.  If $h_R$ is too low, our simulation will yield trees that are too broad and impractical to deal with.  Additionally, our called peaks will be very wide.  If $h_R$ is too high, our simulated trees will be essentially trivial and will not reflect any information on shape.  For a fixed $h$, define $S(h)$ to be the expected length of an excursion above height $h$.  This is the expected return time for state $h$ in the Markov chain $Z$ with transition probabilities $P(i,j)$ from Equation \eqref{markov}.  We calculate $S(h)$ using the fact $S(h)=\frac{1}{\pi(h)}$, where $\pi$ is the stationary distribution given by the recurrence
\begin{equation}
\pi(i) =
\begin{cases}
\pi(h+1) P(h+1,h),& \quad \text{if $i=h$}, \\
\pi(i-1) P(i-1,i) + \pi(i+1) P(i+1,i),& \quad \text{if $i >h$}. \\
\end{cases}
\end{equation}
These equations have a unique solution up to a multiplicative constant,
and that constant is determined by the condition that $\sum_{i}\pi(i) =1$ \cite[\textsection 6.4]{Grimmett-book}.  If $\theta$ is the expected height of the coverage function on $R$, then we define $$h_R:=\max{\left(\lceil \theta \rceil,   \min_{S(h)\leq C}(h)\right)},$$ where $C$ is a parameter that keeps the simulated trees to a reasonable size.  Due to the great variability in the return times for $Z$, we use $C=7$ in our analysis.

 \subsection{Correcting for multiple hypotheses}
 With $\alpha=0.01$ as the significance level, we use a Benjamini--Hochberg correction \cite{Bejamini, Benjamini2} for multiple hypothesis testing.  Once we have assigned probabilities to the $N$ subtrees found on the entire genome, we order the probabilities from least to greatest $p_{(1)}\le p_{(2)} \le \ldots \le p_{(N)}$.  Let $J$ be the largest $j$ such that $p_{(j)} \le \frac{j\alpha}{ N}$.   Then, a tree $T$ in a region $R$ is a called peak provided $G_{R}(\mathcal{M}((T))\le \frac{J \alpha}{N}$.
\section{Results}
\subsection{Predicting STAT1 binding sites as compared to PeakSeq}
\begin{figure}[!ht]
\begin{center}
\includegraphics[width=4.7in]{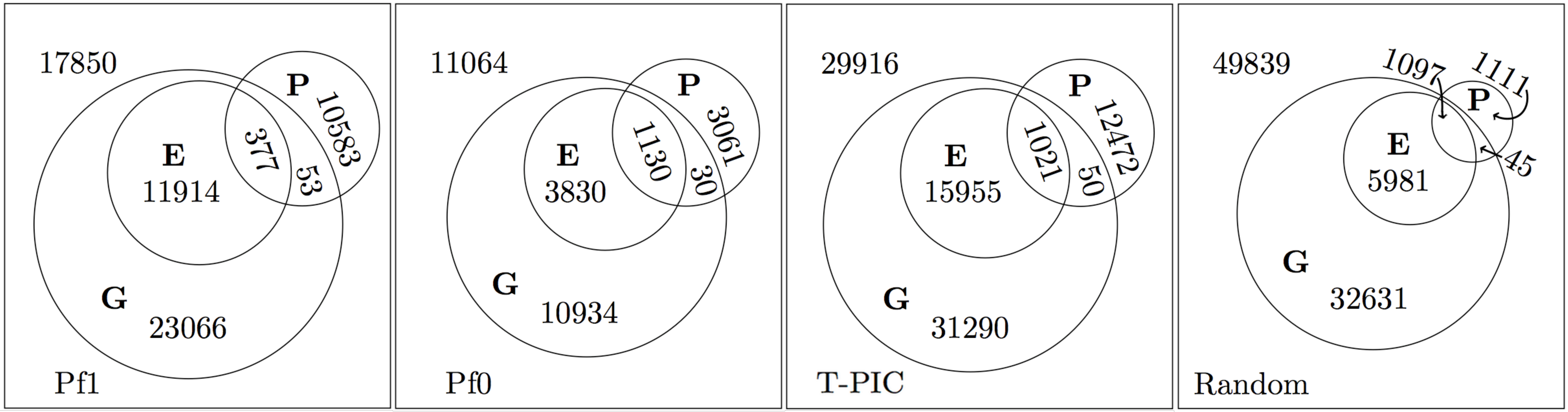}
\caption{The relationships between genomic locations of peaks called by T-PIC and by PeakSeq with normalization parameters $P_f=0$ and $P_f=1$ are depicted with Venn Diagrams.  We additionally show the Venn Diagram for random sequences having the same number and length from each chromosome as our called peaks but with a random start site.  The circular regions include peaks overlapping genes (G), overlapping exons (E), and near promoter (P).}
\label{STAT1Venn}
\end{center}
\end{figure}

Our method predicts 90,704 binding sites for STAT1 on the human genome using the Gerstein data \cite{PeakSeqCode}.  We additionally used PeakSeq's publicly available code \cite{PeakSeqCode} with the default parameters to predict binding sites for the same data.  PeakSeq's method involves a two-pass approach \cite{PeakSeq}.  Initially, PeakSeq finds 122,348 potential binding sites (66.8\% of these overlap our peaks).  In the second pass, PeakSeq identifies the significant peaks based on a parameter $P_f$ with $0\leq P_f \leq 1$.  The most conservative choice is $P_f=0$ in which case PeakSeq calls 30,049 binding sites (93.8\% overlap our peaks).  On the other extreme, using $P_f=1$ yields the most liberal peak calling with 63,843 binding sites (86.2\% overlap with our peaks).  Of the 36,743 of our peaks that PeakSeq did not find with either normalization, 14.1\% (resp. 6.9\%) are within 1000 bp of one of PeakSeq's called peaks for $P_f=1$ (resp. $P_f=0$).   
 \begin{table}
\caption{Comparison with PeakSeq \cite{PeakSeq} and Gerstein's STAT1 data}
{\begin{tabular}{lllllll} \hline\noalign{\smallskip}
&Mean&\# of & \% Near & \% Overlapping &\% Overlapping & \% Containing \\&Length
&Peaks &Promoter$^\dagger$& Exon& Gene & TTCNNNGAA$^{\ddagger}$ \\ \noalign{\smallskip} 
\hline 
\noalign{\smallskip} 
First Pass &363.4&122,348& 12.3&14.2&53.9&17.2 \\
 $P_f=1$&550.4 &63,843& 17.3 &19.3&53.0& 27.7 \\
$P_f=0$&642.9&30,049 & 14.0 &16.5& 55.5&42.2 \\ 
T-PIC&807.7& 90,704& \bf{14.9} &\bf{18.7}& \bf{53.3} & \bf{31.1}\\
Random &807.7 &90,704 &2.5& 7.8&43.8  &22.3 \\ \hline
\end{tabular}}{ \\ $^\dagger$ `Near promoter' is within 1,000 bp upstream and 200 bp downstream as in \cite{Akshay2007}\\ $^{\ddagger}$binding motif for STAT1 \cite{Akshay2007} }
\label{PeakSeqTable}
\end{table}
Table \ref{PeakSeqTable} and Figure \ref{STAT1Venn} compare our predicted binding sites with PeakSeq's in terms of the genomic location of peaks.  We also compare our predictions to random sequences.  These are created by taking a random start site in the same chromosome and the same length for each of our called peaks.  Additionally, Table \ref{PeakSeqTable} shows the number of peaks containing STAT1's binding motif.  Even though we predict more significant peaks than PeakSeq does, we outperform PeakSeq (for at least one choice of $P_f$) for the percentages in Table \ref{PeakSeqTable} shown in bold.  This suggests that our additional peaks contain true binding sites overlooked by PeakSeq's method.  
 
\begin{figure}[!ht]
\begin{center}
\includegraphics[width=4.7in]{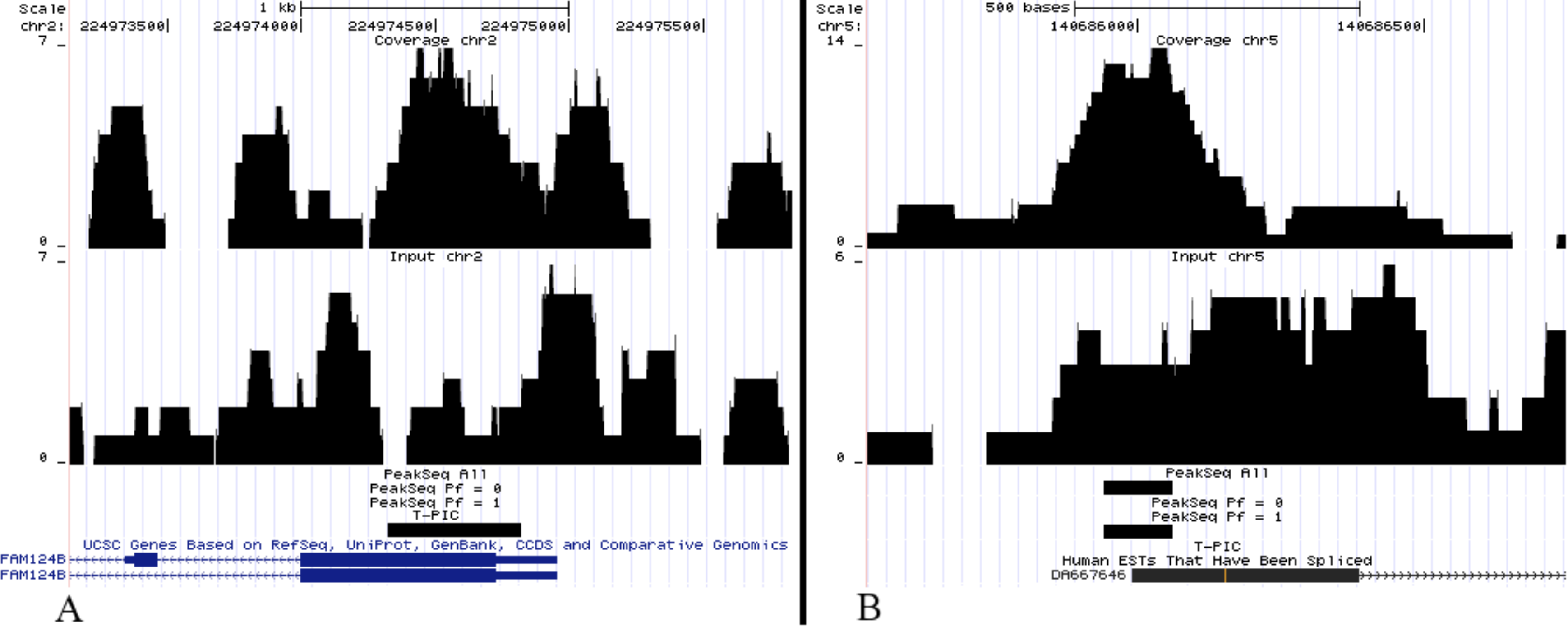}
\caption{Two examples that illustrate differences between peaks called by T-PIC and those predicted by PeakSeq.  The coverage function for both the data and the input control are plotted for each example.  We call the peak in (A) which is near a promoter and not found by PeakSeq.  PeakSeq's significant peak (B) overlaps an expressed sequence tag (EST) but no known gene and was not predicted by T-PIC.}
\label{STAT1UCSC}
\end{center}
\end{figure}

Figure \ref{STAT1UCSC} gives two examples of how our called peaks differ from those predicted by PeakSeq.  One example (Figure \ref{STAT1UCSC}A) contains one of our peaks that lies near a promoter.  PeakSeq's initial pass through the data did not select this peak of height seven for further analysis.  Indeed, PeakSeq's height cutoff for this million bp in chromosome 2 was eight.  Figure \ref{STAT1UCSC}B depicts a significant peak for PeakSeq with normalization parameter $P_f=1$.  This peak overlaps an EST but no known genes, and the tree we found on this interval did not have a statistically significant shape. 

\subsection{Predicting binding sites for Drosophila Melanogaster as compared to MACS}
With our algorithm, we predict binding sites for six transcription factors (with a total of 8 antibodies) for Drosophila Melanogaster.  We use published data from the Eisen lab \cite{bradley} (available at the NCBI GEO database \cite{Barrett2009}, accession GSE20369) and compare our method to MACS \cite{MACS}.  Peaks are called with MACS using the parameters from Bradley et al.'s original study \cite{bradley}.  Tables \ref{MACSTable} and \ref{TreeTable} give summaries of the peaks called by MACS and T-PIC.  
\begin{table}[!ht]
\caption{MACS predictions for Eisen's Drosophila melanogaster data.}
{\begin{tabular}{lllllll} \hline\noalign{\smallskip} 
Protein & Mean& \# of&\% Found &\% Near&\% Overlapping  &\% Overlapping \\
&Length&Peaks& By Our Method &Promoter$^\dagger$& Exon & Gene \\ \noalign{\smallskip} 
\hline 
\noalign{\smallskip} 
bcd&1,639&2,048 & 99&42.3&59.2 &75.4 \\
cad&1,594& 4,271&96.6&50.3&55 &75.2\\
gt& 1,172.9& 2,788&90.7 &27.9& 31.4&60.8\\ 
hb1 &1,832.5& 4,715&97.5 &45.8&51.4&74.1\\
hb2& 1,449.7&4,415&95.4 &42.2&46.8&72\\ 
kni& 1,983.9& 535&85.2&40&44.7&64.9\\
kr1& 1,529.7&6,391&99.5 &35.9&42.1&66\\
kr2& 1,473.1&6,160&98.6&34.5&41.1&65.7\\
\hline
\end{tabular}}{ \\$^\dagger$ `Near promoter' is within 1,000 bp upstream and 200 bp downstream as in \cite{Akshay2007}}

\label{MACSTable}
\end{table}

\begin{figure}
\begin{center}
\includegraphics[width=4.7in]{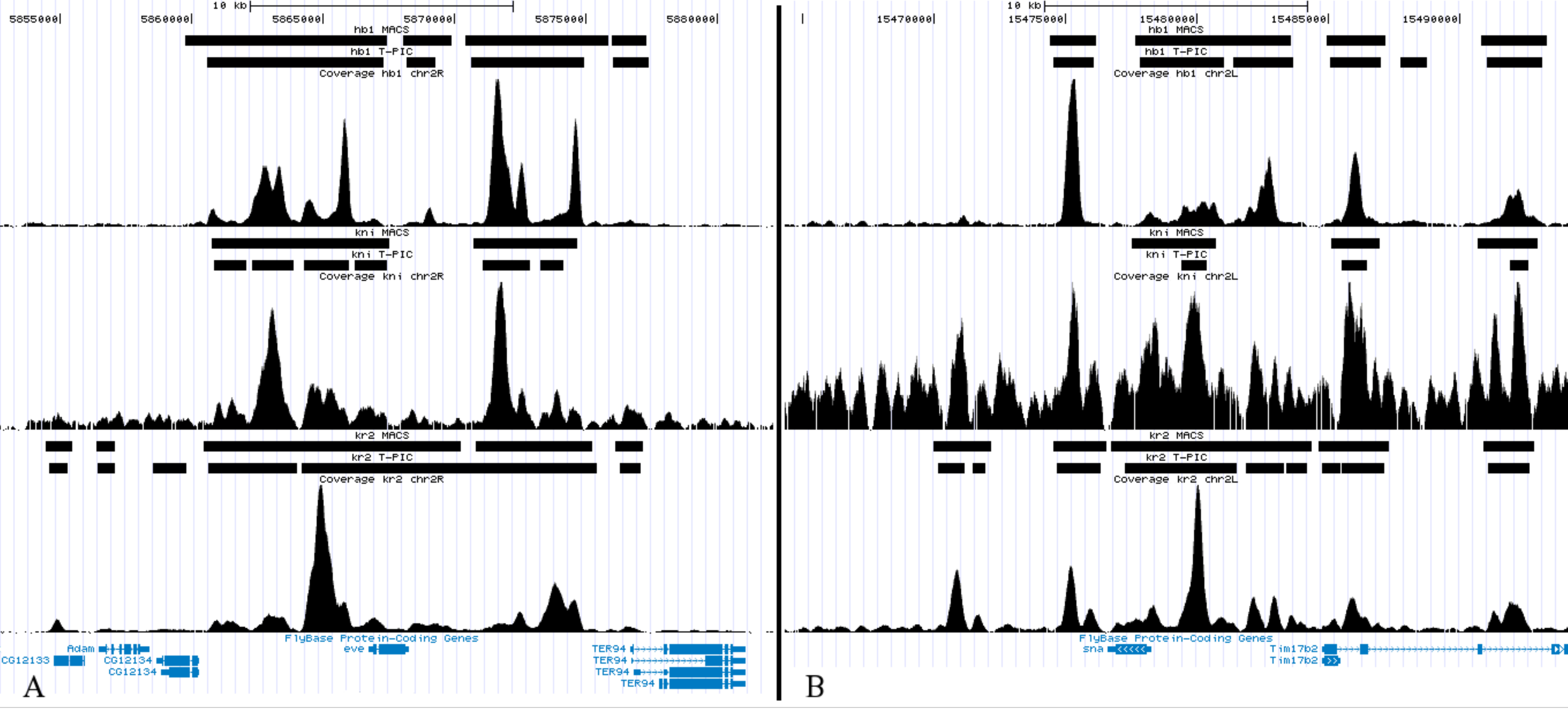}
\caption{ Peaks called by both MACS and T-PIC for thee transcription factors---hunchback antibody 1 (hb1), knirps (kni), and kruppel antibody 2 (kr2)---in the even skipped (A) and snail (B) loci are shown.  The plotted coverage function for each transcription factor suggests multiple peaks, and our method better ``sees'' the multiple binding sites.}
\label{Mel-Loci}
\end{center}
\end{figure}

\begin{table}
\caption{T-PIC predictions for Eisen's Drosphila melanogaster data.}
{\begin{tabular}{llllllll}\hline\noalign{\smallskip} 

Protein & Mean& \# of& \# Not Found &\# w/i 1000 bp&\% Near&\% Overlapping  &\% Overlapping \\
&Length&Peaks& by MACS&of MACS peak&Promoter$^\dagger$& Exon & Gene \\ \noalign{\smallskip} 
\hline 
\noalign{\smallskip} 
bcd&661.9&16,262&12,591&498&22.2&45.2& 64.8\\
cad&942&8,231&3,019&336&36.7&44.5& 66.1\\
gt& 881.4&4,774&1,831&123&21.5&26.1&57.7\\ 
hb1 &981.3&7,497 &1,867&229&35.2&38.0&65.7\\
hb2&923.6&6,528&1,675&221&34.0&37.1&65.2 \\ 
kni&974.3&2,103&1,270&14&22.6&31.1& 54.4\\
kr1&844.7&12,328 &4,223&727&27.7&33.9&61.3\\
kr2& 883.2&11,588&4,049&687&27.5&33.6&60.5\\
\hline
\end{tabular}}{ \\$^\dagger$ `Near promoter' is within 1,000 bp upstream and 200 bp downstream as in \cite{Akshay2007}}

\label{TreeTable}
\end{table}
For all proteins, we call shorter peaks than MACS.  While the percentages of MACS peaks near a promoter and overlapping genes/exons are greater than those for T-PIC, we predict substantially more peaks and a reasonable percentage of them are still located in gene regions.  Moreover, for our called peaks that do not overlap peaks called by MACS, a fair number of them fall within 1000 bp of a MACS peak, as shown in Table \ref{TreeTable}.  Figure \ref{Mel-Loci} shows called peaks for three of the transcription factors in the even skipped (eve) and snail (sna) loci along with the coverage function for each transcription factor.  The binding for these two well-characterized loci has been previously studied \cite{19627575}.  In many cases, our peaks subdivide those called by MACS, which illustrates that our method is more sensitive to the presence of multiple binding sites.  In other cases, we narrow a peak called by MACS and give a more specific location to the DNA binding.  In Figure \ref{Mel-Loci}, we additionally find a few new examples of binding sites that MACS missed.

\section{Conclusion}

We have developed a novel approach to the analysis of ChIP-Seq data, that aims to discover bound regions of DNA by topological analysis of read coverage functions. Our method--T-PIC--can be used with or without an input control, is fast, and freely available, making it suitable for general use.  The approach compares favorably to two popular peak callers: PeakSeq and MACS.  We find the majority of their called peaks while detecting additional sites of binding, improving upon percentages (PeakSeq), and separating peaks into multiple binding sites (MACS). Although we have focused on ChIP-Seq in this paper, the approach we describe to call peaks could also be of use in the analysis of other sequence-based assays. 
\bibliography{bibfile}{}

\begin{thebibliography}{10}
\providecommand{\url}[1]{\texttt{#1}}
\providecommand{\urlprefix}{URL }

\bibitem{Barrett2009}
Barrett, T., Troup, D.B., Wilhite, S.E., Ledoux, P., Rudnev, D., Evangelista,
  C., Kim, I.F., Soboleva, A., Tomashevsky, M., Marshall, K.A., Phillippy,
  K.H., Sherman, P.M., Muertter, R.N., Edgar, R.: {NCBI GEO: archive for
  high-throughput functional genomic data}. Nucl. Acids Res.  37(suppl1),
  D885--890 (2009),
  \url{http://nar.oxfordjournals.org/cgi/content/abstract/37/suppl_1/D885}

\bibitem{Barski2009}
Barski, A., Zhao, K.: Genomic location analysis by {C}h{IP}-{S}eq. Journal of
  Cellular Biochemistry  107(1),  11--18 (2009),
  \url{http://dx.doi.org/10.1002/jcb.22077}

\bibitem{Bejamini}
Benjamini, Y., Hochberg, Y.: Controlling the false discovery rate: a practical
  and powerful approach to multiple testing. J. Roy. Statist. Soc. Ser. B
  57(1),  289--300 (1995),
  \url{http://links.jstor.org/sici?sici=0035-9246(1995)57:1<289:CTFDRA>2.0.CO;%
2-E&origin=MSN}

\bibitem{Benjamini2}
Benjamini, Y., Yekutieli, D.: The control of the false discovery rate in
  multiple testing under dependency. Ann. Statist.  29(4),  1165--1188 (2001),
  \url{http://dx.doi.org/10.1214/aos/1013699998}

\bibitem{bhamidi2009}
Bhamidi, S., Evans, S.N., Sen, A.: Spectra of large random trees (2009),
  \url{http://www.citebase.org/abstract?id=oai:arXiv.org:0903.3589}

\bibitem{Akshay2007}
Bhinge, A.A., Kim, J., Euskirchen, G.M., Snyder, M., Iyer, V.R.: {Mapping the
  chromosomal targets of {STAT}1 by Sequence Tag Analysis of Genomic Enrichment
  (STAGE)}. Genome Research  17(6),  910--916 (2007),
  \url{http://genome.cshlp.org/content/17/6/910.abstract}

\bibitem{Blahnik2010}
Blahnik, K.R., Dou, L., O'Geen, H., McPhillips, T., Xu, X., Cao, A.R., Iyengar,
  S., Nicolet, C.M., Ludascher, B., Korf, I., Farnham, P.J.: {Sole-Search: an
  integrated analysis program for peak detection and functional annotation
  using ChIP-seq data}. Nucl. Acids Res.  38(3),  e13-- (2010),
  \url{http://nar.oxfordjournals.org/cgi/content/abstract/38/3/e13}

\bibitem{bradley}
Bradley, R.K., Li, X.Y., Trapnell, C., Davidson, S., Pachter, L., Chu, H.C.,
  Tonkin, L.A., Biggin, M.D., Eisen, M.B.: Binding site turnover produces
  pervasive quantitative changes in transcription factor binding between
  closely related drosophila species. PLoS Biol  8(3),  e1000343 (03 2010),
  \url{http://dx.doi.org/10.1371%2Fjournal.pbio.1000343}

\bibitem{Carlsson-2009}
Carlsson, G.: Topology and data. Bull. Amer. Math. Soc. (N.S.)  46(2),
  255--308 (2009), \url{http://dx.doi.org/10.1090/S0273-0979-09-01249-X}

\bibitem{EvHoPa}
{Evans}, S.N., {Hower}, V., {Pachter}, L.: {Coverage statistics for sequence
  census methods}. ArXiv e-prints  (Apr 2010),
  \url{http://arxiv.org/abs/1004.5587}

\bibitem{Evans-book}
Evans, S.N.: Probability and real trees, Lecture Notes in Mathematics, vol.
  1920. Springer, Berlin (2008), lectures from the 35th Summer School on
  Probability Theory held in Saint-Flour, July 6--23, 2005

\bibitem{MR0400434}
Fearn, D.H.: Galton-{W}atson processes with generation dependence. In:
  Proceedings of the {S}ixth {B}erkeley {S}ymposium on {M}athematical
  {S}tatistics and {P}robability ({U}niv. {C}alifornia, {B}erkeley, {C}alif.,
  1970/1971), {V}ol. {IV}: {B}iology and health. pp. 159--172. Univ. California
  Press, Berkeley, Calif. (1972)

\bibitem{Fejes2008}
Fejes, A.P., Robertson, G., Bilenky, M., Varhol, R., Bainbridge, M., Jones,
  S.J.M.: {FindPeaks 3.1: a tool for identifying areas of enrichment from
  massively parallel short-read sequencing technology}. Bioinformatics  24(15),
   1729--1730 (2008),
  \url{http://bioinformatics.oxfordjournals.org/cgi/content/abstract/24/15/172%
9}

\bibitem{MR0065835}
Good, I.J.: The joint distribution for the sizes of the generations in a
  cascade process. Proc. Cambridge Philos. Soc.  51,  240--242 (1955)

\bibitem{Grimmett-book}
Grimmett, G.R., Stirzaker, D.R.: Probability and random processes. Oxford
  University Press, New York, third edn. (2001)

\bibitem{MR1991122}
Harris, T.E.: The theory of branching processes. Dover Phoenix Editions, Dover
  Publications Inc., Mineola, NY (2002), corrected reprint of the 1963 original
  [Springer, Berlin; MR0163361 (29 \#664)]

\bibitem{Ji2008}
Ji, H., Jiang, H., Ma, W., Johnson, D.S., Myers, R.M., Wong, W.H.: An
  integrated software system for analyzing {C}h{IP}-chip and {C}h{IP}-seq data.
  Nat Biotech  26(11),  1293--1300 (11 2008),
  \url{http://dx.doi.org/10.1038/nbt.1505}

\bibitem{Johnson2007}
Johnson, D.S., Mortazavi, A., Myers, R.M., Wold, B.: {Genome-Wide Mapping of in
  Vivo Protein-DNA Interactions}. Science  316(5830),  1497--1502 (2007),
  \url{http://www.sciencemag.org/cgi/content/abstract/316/5830/1497}

\bibitem{Jothi2008}
Jothi, R., Cuddapah, S., Barski, A., Cui, K., Zhao, K.: {Genome-wide
  identification of in vivo protein-DNA binding sites from {C}h{IP}-{S}eq
  data}. Nucl. Acids Res.  36(16),  5221--5231 (2008),
  \url{http://nar.oxfordjournals.org/cgi/content/abstract/36/16/5221}

\bibitem{Kharchenko2008}
Kharchenko, P.V., Tolstorukov, M.Y., Park, P.J.: Design and analysis of
  {C}h{IP}-seq experiments for dna-binding proteins. Nat Biotech  26(12),
  1351--1359 (12 2008), \url{http://dx.doi.org/10.1038/nbt.1508}

\bibitem{19627575}
MacArthur, S., Li, X.Y., Li, J., Brown, J., Chu, H.C., Zeng, L., Grondona, B.,
  Hechmer, A., Simirenko, L., Keranen, S., Knowles, D., Stapleton, M., Bickel,
  P., Biggin, M., Eisen, M.: Developmental roles of 21 {D}rosophila
  transcription factors are determined by quantitative differences in binding
  to an overlapping set of thousands of genomic regions. Genome Biology  10(7),
   R80 (2009), \url{http://genomebiology.com/2009/10/7/R80}

\bibitem{Nix2008}
Nix, D., Courdy, S., Boucher, K.: Empirical methods for controlling false
  positives and estimating confidence in {C}h{IP}-{S}eq peaks. BMC
  Bioinformatics  9(1),  523 (2008),
  \url{http://www.biomedcentral.com/1471-2105/9/523}

\bibitem{Park2009}
Park, P.J.: {C}h{IP}-seq: advantages and challenges of a maturing technology.
  Nat Rev Genet  10(10),  669--680 (2009),
  \url{http://dx.doi.org/10.1038/nrg2641}

\bibitem{Pepke2009}
Pepke, S., Wold, B., Mortazavi, A.: Computation for {C}h{IP}-seq and {RNA}-seq
  studies. Nat Meth  6(11s),  S22--S32 (11 2009),
  \url{http://dx.doi.org/10.1038/nmeth.1371}

\bibitem{Robertson2007}
Robertson, G., Hirst, M., Bainbridge, M., Bilenky, M., Zhao, Y., Zeng, T.,
  Euskirchen, G., Bernier, B., Varhol, R., Delaney, A., Thiessen, N., Griffith,
  O.L., He, A., Marra, M., Snyder, M., Jones, S.: Genome-wide profiles of
  {STAT}1 {DNA} association using chromatin immunoprecipitation and massively
  parallel sequencing. Nat Meth  4(8),  651--657 (08 2007),
  \url{http://dx.doi.org/10.1038/nmeth1068}

\bibitem{PeakSeqCode}
Rozowsky, J., Euskirchen, G., Auerbach, R.K., Zhang, Z.D., Gibson, T.,
  Bjornson, R., Carriero, N., Snyder, M., Gerstein, M.B.: Supplemental code and
  data for {P}eak{S}eq: scoring {C}h{IP}-seq experiments relative to controls,
  \texttt{http://www.gersteinlab.org/proj/PeakSeq/}

\bibitem{PeakSeq}
Rozowsky, J., Euskirchen, G., Auerbach, R.K., Zhang, Z.D., Gibson, T.,
  Bjornson, R., Carriero, N., Snyder, M., Gerstein, M.B.: Peak{S}eq enables
  systematic scoring of {C}h{IP}-{S}eq experiments relative to controls. Nat
  Biotech  27,  66--75 (2009), \url{http://dx.doi.org/10.1038/nbt.1518}

\bibitem{Wold2008}
Wold, B., Myers, R.M.: Sequence census methods for functional genomics. Nat
  Meth  5(1),  19--21 (2008), \url{http://dx.doi.org/10.1038/nmeth1157}

\bibitem{MACS}
Zhang, Y., Liu, T., Meyer, C., Eeckhoute, J., Johnson, D., Bernstein, B.,
  Nussbaum, C., Myers, R., Brown, M., Li, W., Liu, X.S.: Model-based analysis
  of {C}h{IP}-{S}eq ({MACS}). Genome Biology  9(9),  R137 (2008),
  \url{http://genomebiology.com/2008/9/9/R137}

\end{thebibliography}
\bibliographystyle{splncs}

 \end{document}